\documentclass[10pt]{article}
\usepackage{latexsym}

\newcommand{\np}{Nucl.\,Phys.\,}
\newcommand{\pl}{Phys.\,Lett.\,}
\newcommand{\pr}{Phys.\,Rev.\,}

\newcommand{\zp}{Z.\,Phys.\,}

\def\ds {\displaystyle}
 
\def\expf{\mathrel{\mathop{e}
\limits_{\raisebox{1.5mm}{\scriptsize $\rightarrow$}}}} 
\def\expb{\mathrel{\mathop{e}
\limits_{\raisebox{1.5mm}{\scriptsize $\leftarrow$}}}}
\newcommand{\be}{\begin{equation}}
\newcommand{\ee}{\end{equation}}
\newcommand{\bea}{\begin{eqnarray}}
\newcommand{\ena}{\end{eqnarray}} 
\newcommand{\vp}{\varphi} 
\newcommand{\eps}{\varepsilon}

\newcommand{\tr}{{\rm Tr}}

\begin{document}

%%%%%%%%%%%%%%%%%%%%%%%%%%%%%%%%%%%%%%%%%%
%%%%%%%%%%%% Biblio style %%%%%%%%%%%%%%%%
%%%%%%%%%%%%%%%%%%%%%%%%%%%%%%%%%%%%%%%%%%
\bibliographystyle{unsrt}
%%%%%%%%%%%%%%%%%%%%%%%%%%%%%%%%%%%%%%%%%%

%%%%%%%%%%%%%%%%%%%%%%%%%%%%%%%%%%%%%%%%%%%%%%%%%%%%%%%%%%%%%%%%%%%%%%%%
%%%%%%%%%%%%%%%%%%%%%%%%%%%% TITLEPAGE %%%%%%%%%%%%%%%%%%%%%%%%%%%%%%%%% 
%%%%%%%%%%%%%%%%%%%%%%%%%%%%%%%%%%%%%%%%%%%%%%%%%%%%%%%%%%%%%%%%%%%%%%%%
\begin{titlepage}

\def\baselinestretch{1.2}
\vspace*{\fill}
\begin{center}
%%%%%%%%%%%%%%%%%%%%%%%%%%%%%%% TITLE %%%%%%%%%%%%%%%%%%%%%%%%%%%%%%
{\Large {\bf Motion of a scalar field coupled to a Yang-Mills 
field reformulated locally with some gauge invariant variables }}
%%%%%%%%%%%%%%%%%%%%%%%%%%%%%%%%%%%%%%%%%%%%%%%%%%%%%%%%%%%%%%%%%%%%
\vspace*{0.5cm}

 E.~Chopin\\ 

{\small E-mail:~eric.chopin@wanadoo.fr} 
\end{center}
\vspace*{\fill}

\centerline{ {\bf Abstract} }
\baselineskip=14pt
\noindent
%%%%%%%%%%%%%%%%%%%%%%%%%%% ABSTRACT %%%%%%%%%%%%%%%%%%%%%%%%%%%%%%
 {\small This paper exposes a reformulation of some gauge 
theories in terms of explicitly gauge-invariant variables. 
We show in the case of Scalar QED that the classical theory 
can be reformulated locally with some gauge invariant variables. 
We discuss the form of some 
realistic asymptotic solutions to these equations. The 
equations of motion are then also reformulated in the non-abelian 
case. 
}
%%%%%%%%%%%%%%%%%%%%%%%%%%%%%%%%%%%%%%%%%%%%%%%%%%%%%%%%%%%%%%%%%%%%%
\vspace*{\fill}
 
%\rightline{hep-ph/0001125}
%\rightline{Jan. 2000}
\rightline{Report No:THP/00/001}
\rightline{Journal ref:JHEP 03 (2000) 022}

\end{titlepage}
\baselineskip=14pt
\def\baselinestretch{1.0}

%%%%%%%%%%%%%%%%%%%%%%%%%%%%%%%%%%%%%%%%%%%%%%%%%%%%%%%%%%%%%%%%%%%%%
%%%%%%%%%%%%%%%%%%%%%%%%%%% INTRODUCTION %%%%%%%%%%%%%%%%%%%%%%%%%%%%
%%%%%%%%%%%%%%%%%%%%%%%%%%%%%%%%%%%%%%%%%%%%%%%%%%%%%%%%%%%%%%%%%%%%%

\section*{Introduction}

The gauge symmetry is known to render the calculation 
of the elements of the S matrix very intricate. 
In some future colliders like LHC or NLC, 
some very complicated scattering processes will be studied. 
 Phenomenologists will 
have to consider processes with 3, 4 or more 
particles in the final state. The scattering amplitudes for 
these processes are in general very complicated because 
of the very large number of Feynman graphs, and the numerical 
evaluation of these amplitudes in Monte-Carlo programs 
suffer from numerical instabilities due for a large 
part to some huge compensations between the different 
graphs, which arise from the gauge symmetry. 
To avoid these numerical instabilities, there are two common methods.
 The first one consists in using a specific gauge which 
simplifies the different vertices and 
propagators~\cite{georges,fujikawa2,hhh,bfm.other,bckgdenner}. The 
second one consists in using some algorithms acting on each 
Feynman graph, based on Ward identities, in order to simplify the 
expression of the graphs~\cite{pinch}.  Both methods lead to the 
elimination of most of these huge compensations. \\ 

In this paper, we consider this problem 
from another point of view, at the core of 
 Quantum Field Theory. Basically, we raise the question
 of whether the calculations of the elements of the S~matrix can be done 
directly using some gauge invariant variables. This question can 
be studied  in the context of both methods cited above, using as a 
fundamental tool the Ward identities. These identities depend on 
the gauge fixing procedure used in the calculations. We rather 
look here for a method in which
there is no need to break temporarily the gauge symmetry.
 As a consequence, we must start our 
formulation from the very beginning of gauge theories, that is to 
say from the equations of motion. We therefore show in this paper 
that one can reformulate these equations in terms of 
local gauge invariant variables for the case where matter fields 
are scalar. 
\\ 

This new approach may have some interesting applications 
regarding the quantization of fields. In standard 
field theory, the quantization procedure is done first on free fields, 
and therefore matter fields and gauge fields are considered 
separately, though they are coupled in the equations of motion. 
A significant consequence is that it is irrelevant to consider 
the evolution of a free field from a time $t$ to an 
interacting field at time $t'>t$ through a unitary transformation, 
because Haag's theorem says that the field considered at time $t'$ 
must be also free (for a good review, see~\cite{bogoliubov}). 
Quantum Field Theory is therefore doomed to describe 
only the transition between asymptotic fields through the 
LSZ formalism. In experiments where the time variable 
plays a fundamental role (CP violation experiments in $K^0_S/K^0_L$, 
neutrino oscillations,...) one must use a mixed theory, 
based in part on classical quantum mechanics (Rabi precession,...) 
and in part on quantum field theory for the computation of the decay 
width of the particles. A single theory which would 
describe completely such experiments is still missing. Since 
Haag's theorem does not apply to the case of two constantly 
interacting fields, the approach presented in this paper opens  
the prospect of finding an evolution operator between two 
finite times for an 
interacting system. That is to say, 
an asymptotic electron would be 
described both by its matter field and its surrounding 
electromagnetic field, in some sense. So we must also find some 
``realistic'' asymptotic solutions to the coupled equations of 
motion in replacement of the plane waves that are used in standard 
quantum field theory.  The word ``realistic'' means here that we 
look for solutions that have a finite conserved momentum. We show 
that solitons are not possible in this context (for the $U(1)$ 
case), but we conjecture that some periodic-in-time solutions may 
probably exist. \\     

What is the basic idea of our approach? We know that for a given 
field-strength tensor, one can compute  a corresponding gauge 
field using the basic cohomological formulas that are reviewed in 
the appendix. Some authors have already tried to reformulate the 
Yang-Mills Theory using only the Field-Strength tensor as a basic 
variable in place of the gauge 
field~\cite{halpern.fst,seo-okawa.fst,itabashi.fst,mendel.fst}. 
The results of these studies are generally not covariant and 
non-local, due to the fact that the cohomological formulas are 
essentially of a non-local nature. In this paper, we rather 
consider the gauge-currents as fundamental variables, and we keep 
both locality and  covariance of the equations. \\ 
   
The paper is therefore organized as follows: \\ 

The first section is devoted to the reformulation 
of $U(1)$ scalar QED in terms of gauge invariant variables. \\ 

The second section 
contains  a discussion on asymptotic solutions of the $U(1)$ 
scalar QED. We first show that periodic solutions of Klein Gordon 
do not have a finite energy, contrary to what is claimed in a 
recent paper, and therefore we need to consider the coupled 
equations. We show the impossibility of 
soliton solutions  in this context, and discuss the possibility of 
periodic (in time) solutions. \\ 

The third section presents the non-abelian case, where 
the gauge group is in a certain class of subgroups of $U(N)$. 
It turns out that the results presented in this paper 
are in fact a simpler version 
of some results given by Lunev in 1994~\cite{lunev-gauge}, 
with in addition 
the coupling to a scalar multiplet (he only considered 
a pure Yang-Mills theory). \\ 

%%%%%%%%%%%%%%%%%%%%%%%%%%%%%%%%%%%%%%%%%%%%%%%%%%%%%%%%%%%%%%%%%
%%%%%%%%%%%%%%%%%%% SECTION 1 %%%%%%%%%%%%%%%%%%%%%%%%%%%%%%%%%%%
%%%%%%%%%%%%%%%%%%%%%%%%%%%%%%%%%%%%%%%%%%%%%%%%%%%%%%%%%%%%%%%%%

\section{A possible reformulation of classical SQED}
\label{sqed} 

In this section, we reformulate the classical theory of 
a scalar field coupled to a $U(1)$ gauge field (SQED) 
in terms of gauge invariant variables. We will then 
demonstrate
  the difficulties appearing when one wants to find 
some ``realistic'' asymptotic solutions, which would generate a 
Fock-like space. Using such a space, one could then construct a 
new formalism for computing cross sections. 
 Let us start with the classical scalar QED lagrangian:

\be
{\cal L} = (D_\mu\phi)^* D^\mu \phi -m^2 \phi^*\phi -\frac{1}{2}
\partial_\mu A_\nu (\partial^\mu A^\nu - \partial^\nu A^\mu) 
\label{sqed-lagr}
\ee

 with $D_\mu = \partial_\mu + ieA_\mu$. The electrical current is 
given by $J_\mu = ie(\phi^*(D_\mu\phi) - (D_\mu\phi)^*\phi)$ 
 and the probability density $\rho = \phi^*\phi$. Both 
$J_\mu$ and $\rho$ are gauge invariant variables and we will show 
how to rewrite the previous lagrangian as a function of these 
variables (this treatment will have to be modified in the 
non-abelian case in which the corresponding expression for these 
variables are not gauge invariant but gauge ``covariant''). First, 
we shall review the standard equations of motion when $\phi$ and 
$A^\mu$ are taken as field variables: 

\bea 
0 &=& (D_\mu D^\mu+m^2)\phi \label{scalar0} \\ 
0 &=& (\Box+m^2)\phi 
+ 2ieA_\mu\partial^\mu \phi 
+ie(\partial\cdot A)\phi -e^2(A\cdot A)\phi \label{scalar} \\
\partial^\alpha F_{\alpha\beta} &=& ie(\phi^*(D_\beta\phi) - 
(D_\beta\phi)^*\phi ) = J_\beta \label{gf} 
\ena

We shall first note that if one computes $\phi^*(\ref{scalar}) - 
(\ref{scalar})^*\phi$, one obtains $\partial_\mu J^\mu =0$, which 
we would have already obtained by taking the divergence of 
eq.~\ref{gf}. The redundancy between the last two equations can 
therefore be removed by making use of $\phi^*(\ref{scalar}) +  
(\ref{scalar})^*\phi$ instead of Eq.~\ref{scalar}. After some 
algebra, it is not a hard task to make $J_\mu$ and $\rho$ appear 
in the equations as we will see later, but for the derivation of 
the new equations, we rather choose to start from the lagrangian. 
For this purpose, we will use the following relations:  

\bea
-\frac{J^2}{e^2} &=& (\phi^*(D_\mu\phi) + (D_\mu\phi)^*\phi)^2 
-4 (D_\mu\phi)^*\phi\phi^*(D_\mu\phi) \label{jsquarre} \\ 
\Rightarrow (D_\mu\phi)^* D^\mu \phi &=& \frac{1}{4\rho} 
\left( (\partial_\mu\rho)^2 + \frac{J^2}{e^2} \right) \label{jsquarre2} \\ 
&=& (\partial_\mu\sqrt{\rho})^2 + \frac{J^2}{4e^2\rho} \label{jsq2} 
\ena 

 Throughout the paper, we will conveniently 
define $v^\mu$ such that $J^\mu = 2e^2 \rho v^\mu$ 
and set $z(x) = \sqrt{\rho(x)}$. 
From the definition of the current, one can also extract 
the expression of the field strength tensor: 

\bea
2e^2F_{\mu\nu} &=& \partial_\mu\left(\frac{J_\nu}{\rho}\right) 
- \partial_\nu\left(\frac{J_\mu}{\rho}\right) \label{ym-from-c}  \\ 
F_{\mu\nu} &=& \partial_\mu(v_\nu) - \partial_\nu(v_\mu) 
\label{def-fmunu}
\ena

Equations~\ref{jsq2} and~\ref{def-fmunu} are the fundamental tools 
of our formalism. With these, we can write the lagrangian as a 
function of $z$ and $v^\nu$ in the following way: 

\be
{\cal L} = (\partial_\mu z)^2 -m^2 z^2 + e^2 z^2 v^2 
-\frac{1}{4}(\partial_\mu(v_\nu) - \partial_\nu(v_\mu))^2 
\label{new-lagr} 
\ee

We have now re-expressed the lagrangian in terms 
of gauge invariant quantities, and 
as a by-product the  ``effective'' coupling constant is $e^2 = 
4\pi\alpha$ instead of $e$. This means that the sign of $e$ is not 
relevant. Although this does not mean that in a perturbative 
expansion of some solutions, the relevant expansion parameter is 
necessarily $e^2$, it may also be $\sqrt{4\pi\alpha}$ or $|e|$. 
From this new lagrangian we can derive the following equations of 
motion thanks to the Euler-Lagrange equations: 

\bea
(\Box+m^2)z &=&  4\pi\alpha zv^2 \label{sqed-gauge-invar1} \\ 
\Box(v_\nu) -\partial_\nu (\partial\cdot v) &=& 
- 8\pi\alpha z^2 v_\nu \label{sqed-gauge-invar2}
\ena

\subsection{The Energy-Momentum Tensor}

We will further look for asymptotic solutions to the 
coupled equations with a finite conserved momentum. 
The symmetrized energy momentum tensor (or Belinfante tensor) 
can be rewritten this way: 

\bea
T_{\mu\nu} &\stackrel{def}{=}& (D_\mu\phi)^\dagger D_\nu\phi +
(D_\nu\phi)^\dagger D_\mu\phi -F_{\mu\lambda}F_\nu^{~\lambda} 
-g_{\mu\nu}\left((D_\mu\phi)^\dagger D^\mu\phi
-m^2\phi^\dagger\phi-\frac{1}{4}F_{\alpha\beta}F^{\alpha\beta} 
\right)\label{em-tensor} \\ 
T_{\mu\nu}&=& 2e^2z^2v_\mu v_\nu +2\partial_\mu z 
\partial_\nu z -\left(\partial_\mu (v_\lambda) 
-\partial_\lambda (v_\mu)\right) 
 \left(\partial_\nu (v^\lambda) 
-\partial^\lambda (v_\nu)\right) \nonumber \\ 
&& -g_{\mu\nu}\left((\partial_\mu z)^2 -m^2 z^2 + e^2 z^2 v^2 
-\frac{1}{4}(\partial_\mu(v_\nu) - \partial_\nu(v_\mu))^2
\right)\label{em-tensor2} \\ 
P_\mu &=& \int_\Sigma d\sigma^\nu T_{\nu\mu} \label{cons-momentum} 
\ena

In Eq.~\ref{cons-momentum}, $\Sigma$ represents any space-like 
hyper-surface in the Minkowsky space time, and 
for the sake of simplicity, we will generally take the $t=0$ 
hypersurface for the computation of $P_\mu$. \\ 

%%%%%%%%%%%%%%%%%%%%%%%%%%%%%%%%%%%%%%%%%%%%%%%%%%%%%%%%%%%%%%%%%
%%%%%%%%%%%%%%%%%%% SECTION 2 %%%%%%%%%%%%%%%%%%%%%%%%%%%%%%%%%%%
%%%%%%%%%%%%%%%%%%%%%%%%%%%%%%%%%%%%%%%%%%%%%%%%%%%%%%%%%%%%%%%%%

\section{Solitons solutions are not normalizable}
\label{solitons}

In general, the spatial extent of the wave 
function of a free particle (obeying the Klein Gordon equation) 
increases in time. Here, we will rather look  
for the possibility to find  
``soliton-like'' solution to the coupled field equations of scalar 
QED (i.e. Eq.~\ref{sqed-gauge-invar1} and 
Eq.~\ref{sqed-gauge-invar2}). First, we will show that for the 
Klein Gordon equation, we can find some simple ``soliton-like'' 
solutions but these solutions are not normalizable (similarly to 
plane waves). We obtain the same result when the interaction is 
taken into account, but the arguments used to reject this case are 
different from the free case. 
For this reason, the free case is also presented, even if it can 
be seen as a particular case of the interacting one.  

\subsection{Generalities about solitons}

We will say that a function $f(x)$ defined on space-time
 is a soliton if we can find a time-like momentum 
$p^\mu$ such that:

\be
p^\mu\partial_\mu f = 0 \label{soliton-def}
\ee

This time-like momentum represents the global momentum of the wave 
which moves without deformation. To see this trivial fact, 
Eq.~\ref{soliton-def} simply means that if we are placed in a 
frame where $p^\mu = (m_0,\vec 0)$, then the shape of the wave 
function does not depend on time ($\partial_0 f=0$). Suppose now 
that at time $t=0$ we look at the shape of the wave function. It 
is reasonable to say  that for an asymptotic solution (supposed to 
describe a free scalar particle)  the probability density is 
spherically symmetric. We can deduce from that that the function 
$f$ is a function of only one variable. To be more specific, let 
us consider the two variables \fbox{$u=(p\cdot x)^2-p^2x^2$} and 
\fbox{$\tau=p\cdot x$}. In the ``rest frame'', where $p^\mu = 
(m_0,\vec 0)$ \footnote{We shall remark that  $p^2=m_0^2$, where 
the mass $m_0$ is {\it a priori} different from the mass $m$ 
appearing in the Klein Gordon equation.}, then:   
  
\bea
u &=& (p\cdot x)^2-p^2x^2 \label{udef} \\ 
&=& (m_0t)^2-m_0^2(t^2-\vec x^2) = m_0^2\vec x^2 \label{ucalc} \\ 
\tau &=& p\cdot x = m_0 t \label{vcalc} 
\ena

A ``spherically symmetric'' scalar function $f$ is therefore 
a function of $u$ and $\tau$ only. We have seen that 
 $u\geq 0$ for any $x$ and we will often write 
\fbox{$y=\sqrt{u}$}. We have by construction  
 $u(x^\mu+\lambda p^\mu) 
= u(x^\mu)$, which means that a function of the variable $u$ is 
invariant under any translation in the $p^\mu$ direction. \\ 
For convenience, we will also use the following notations:  

\bea
\lambda^\alpha &=& p^\alpha(p\cdot x)-p^2x^\alpha 
= \frac{1}{2}\partial^\alpha u(x) \label{lambda} \\ 
\lambda^2 &=& -p^2 u(x) = -m_0^2 u(x) \quad \quad\quad  
 p\cdot\lambda = 0  \label{sl01}  \\  
\partial_\alpha \lambda_\beta &=& p_\alpha p_\beta-m_0^2 
g_{\alpha\beta} = \tau_{\alpha\beta}\label{sl03a} \\ 
\partial^\alpha \lambda_\alpha &=& \tau^\alpha_\alpha = 
 -3m_0^2 \label{sl03b} \\ 
p^\alpha \tau_{\alpha\beta} &=& 0 \quad \quad\quad 
\lambda^\alpha \tau_{\alpha\beta} = -m_0^2 \lambda_\beta 
\label{sl03c}\\ 
\partial_\mu \sqrt{u} &=& \frac{\lambda_\mu}{\sqrt{u}}~~~~~~~~~
\partial^\mu \left(\frac{\lambda_\mu}{\sqrt{u}}\right)=  
-\frac{2m_0^2}{\sqrt{u}} \label{sl03d}\\ 
\ena

And in the rest frame, \fbox{$\lambda^\mu = -m_0^2(0,\vec x)$}. 
Then, if the scalar function $f$ is a ``spherically symmetric'' 
soliton, we have: 

\bea
f(x) &=& g(\tau, u=(p\cdot x)^2-p^2x^2) \label{s02} \\ 
0=p^\mu\partial_\mu f &=& p^\mu(p_\mu\partial_0 g 
+2\lambda_\mu \partial_1 g) \label{s03} = p^2\partial_0 g   
\ena

Therefore $g$ does not depend on $\tau$, but only on $u$. 
We therefore obtain  a covariant formulation of the notion of a 
``spherically symmetric'' soliton.  

\subsection{Periodic solutions to the Klein Gordon equation}

Before we look for some asymptotic solutions to the 
coupled equations, we must explain why we cannot 
have some realistic asymptotic states in the free case. 
Of course finite energy solutions to the 
Klein-Gordon equation exist, consisting in wave-packets 
with square integrable momentum densities. But one 
of the criteria we set in order to define a ``realistic'' 
asymptotic field is that the wave-packet must be ``bounded'' 
in space-like directions. We consider here that 
a constantly spreading wave-packet cannot represent the 
state of a stable free particle.   
We show in this paragraph that soliton solutions 
to the Klein-Gordon equation cannot have a finite energy, 
as a particular case of a stronger result concerning 
periodic-in-time solutions. The free scalar lagrangian is: 

\be
{\cal L}= \partial_\mu\Phi^* \partial^\mu\Phi -m^2 \Phi^*\Phi \label{sl04}
\ee

The energy-momentum tensor and the corresponding conserved 
total momentum are: 

\bea
T_{\mu\nu} &=& \partial_\mu\Phi^*\partial_\nu\Phi+
\partial_\nu\Phi^* \partial_\mu\Phi -{\cal L}g_{\mu\nu} 
\label{sl08} \\ 
P_\nu &=& \int_{\Sigma} d\sigma^\mu T_{\mu\nu} \label{sl09}
\ena

The linearity of the equations of motion allows us to 
expand the field in a Fourier serie: 

\bea 
\phi(t,\vec x) &=& \Sigma_{n=-\infty}^\infty a_n(\vec x)
e^{in\omega t} \label{pKG01} \\ 
&=& \Sigma_{n=-\infty}^\infty a_n(r)
e^{in\omega t} \label{pKG02} \\ 
\ena

We first look for solutions of the form 
$\phi = \exp(i\eta p\cdot x)g(\sqrt{u})$, 
where $\eta$ is a real parameter. We have: 

\bea
\partial^\mu \phi &=& e^{i\eta p\cdot x}\left(i\eta g(\sqrt{u})p^\mu 
+\frac{\lambda^\mu}{\sqrt{u}}g'(\sqrt{u}) \right) \label{pKG03} \\ 
\Box \phi &=& e^{i\eta p\cdot x}\left(-m_0^2\eta^2 g(\sqrt{u}) 
-\frac{2m_0^2}{\sqrt{u}}g'(\sqrt{u})-m_0^2g''(\sqrt{u})\right) 
\nonumber \\ 
&=& -\frac{m_0^2}{y} e^{i\eta p\cdot x}\left(\eta^2 t(y)+t''(y) 
\right)~~~~~~(g(y) = t(y)/y) \label{pKG04} \\ 
0 &=& (\Box+m^2)\Phi \nonumber \\  
\Leftrightarrow 0 &=& t''(y) + t(y)\left(\eta^2
-\frac{m^2}{m_0^2}\right) 
\label{pKG05} \\ 
\Rightarrow t(y) &=& A e^{-\sqrt{\frac{m^2}{m_0^2}-\eta^2}\,y} 
\label{pKG06}
\ena

We have not considered the other solution that increases as 
$y$ (or $r$) increases, because we look for normalized solutions. 
Thus, the general solution is: 

\bea
\phi(x) &=& \frac{1}{y}\sum_{|n|\leq \left[\frac{m}{m_0}\right]} 
A_n e^{in p\cdot x}e^{-\sqrt{\frac{m^2}{m_0^2}-n^2}y} 
\label{pKG07} \\ 
&=& \frac{1}{m_0 r}\sum_{|n|\leq \left[\frac{m}{m_0}\right]} 
 A_n e^{in m_0 t}e^{-\sqrt{m^2-n^2 m_0^2}r} \label{pKG08} 
\ena 

The sum has a finite number of terms because we limit 
ourselves to exponentially 
decreasing terms. It will be clear in the following that the 
oscillating solutions for $|n|> \left[\frac{m}{m_0}\right]$ 
will not provide normalizable solutions. 
Contrary to the claim of Hormuzdiar and Hsu in \cite{breather} 
which considered only the large r behaviour, the solutions are 
not normalizable. This is due to their small $r$ behaviour. This 
can be shown by computing the conserved momentum: 
\\ 

\bea
P_0 &=& \int_{t=0} d\vec x\,\left[2\partial_0\phi^* 
\partial_0\phi-g_{00}(\partial_0\phi^*\partial_0\phi 
-\vec\nabla\phi^*\vec\nabla\phi-m^2\phi^*\phi)\right] \label{normal01} \\ 
&=& \int_{t=0} d\vec x\,\left[\partial_0\phi^* 
\partial_0\phi +\vec\nabla\phi^*\vec\nabla\phi+m^2\phi^*\phi)\right]~~
(\geq 0) \label{normal02} \\ 
&=& 4\pi\int_0^\infty r^2dr\, \frac{1}{r^2}\left[
\left|\sum_n inm_0A_ne^{-\sqrt{m^2-n^2m_0^2}r} \right|^2
+\left|\sum_n A_ne^{-\sqrt{m^2-n^2m_0^2}r}\left(\sqrt{m^2-n^2m_0^2}
+\frac{1}{r}\right)\right|^2 \right. \nonumber \\ 
&&\left. +m^2\left|\sum_n A_ne^{-\sqrt{m^2-n^2m_0^2}r} \right|^2 \right]
\ena  

and the $1/r$ term in the second squared term makes the 
integral divergent. The integral converges if $\sum A_n=0$ 
but the computation on another space-like hypersurface $t=t_0\ne 0$ 
would be still divergent, which is an indication that the computation 
at $t=0$ is meaningless, even if it can be accidentally convergent. \\  

\subsection{Solitons for the coupled SQED equations}

The field $v^\mu$ may also be written in a simple 
generic form if we suppose that it  
obeys the spherically-symmetric soliton condition. 
The most general form compatible with the symmetries  
of the solution is given by: 

\bea 
v^\mu(x) &=& a(u)\lambda^\mu + b(u)p^\mu \label{vsoliton} \\ 
v^2 &=& m_0^2(b^2-a^2 u) \label{vdeux}
\ena  

The first term of $v^\mu$ does not contribute to the 
field strength tensor because if we set $A=\int_0^u a(s)ds$, 
then $a(u)\lambda^\mu = \partial^\mu(A(u)/2)$ 
which is a pure gauge term. And we will further demonstrate 
that this term must vanish. However, we will see in the 
next sections that for periodic solutions, this term is 
important. \\ 
We will also need to comply with the classical asymptotic 
conditions at infinity in space-like directions. One 
must therefore have $A^\mu$ decreasing as $1/r$ at infinity, and 
thus $b(u)\simeq C/\sqrt{u}$ when $u\rightarrow \infty$. \\ 

Then we can substitute $v^\mu$ and \fbox{$z(x)=f(u)$} in the equations 
of motion Eq.~\ref{sqed-gauge-invar1} and 
Eq.~\ref{sqed-gauge-invar2}~: 

\bea
(\Box+(m^2-e^2v^2))z &=& 0 \label{sqed-gauge-invar1b} \\ 
\Box(v_\nu) -\partial_\nu (\partial\cdot v) &=& 
- 2e^2 z^2 v_\nu \label{sqed-gauge-invar2b}
\ena

Using the parameterization of $v^\mu$ given in Eq.~\ref{vsoliton} 
one gets: 

\bea
\partial_\alpha v_\beta &=& 
 2\frac{d(a)}{du}\lambda_\alpha\lambda_\beta 
+2\frac{d(b)}{du}\lambda_\alpha p_\beta+ a\tau_{\alpha\beta} \\ 
F_{\mu\nu} = \partial_\mu v_\nu -\partial_\nu v_\mu 
&=& 2 \frac{d(b)}{du}(\lambda\wedge p)^{\mu\nu} = -2m_0^2 \frac{d(b)}{du} 
(x\wedge p)^{\mu\nu} \\ 
F_{\mu\lambda}F_\nu^{~\lambda} &=& 4m_0^4 {\frac{d(b)}{du}}^2 
(x_\mu x_\nu m_0^2 -(p\cdot x)(p_\mu x_\nu+p_\nu x_\mu)
+x^2 p_\mu p_\nu) \\ 
F^2 &=& -8m_0^4{\frac{d(b)}{du}}^2 u \\ 
\Box v_\nu -\partial_\nu(\partial\cdot v)  
 &=& -4m_0^2u\frac{d^2(b)}{du^2}p_\nu
 -6m_0^2\frac{d(b)}{du}p_\nu
\ena

Thus Eq.~\ref{sqed-gauge-invar2b} yields: 
\bea 
 -4m_0^2u\frac{d^2(b)}{du^2}p_\nu
 -6m_0^2\frac{d(b)}{du}p_\nu 
&=& -2e^2f^2(a\lambda_\nu+bp_\nu) \\ 
\Rightarrow~~~~ a&=& 0 \\ 
and~~~~~~~ 4u\frac{d^2(b)}{du^2}
+ 6\frac{d(b)}{du} 
&=& \frac{2e^2}{m_0^2}f^2b \\ 
\fbox{$\ds b(u) = \frac{\tilde b(\sqrt{u})}{\sqrt{u}}$} 
\Rightarrow~~~~ \tilde b'' &=& 
\frac{2e^2}{m_0^2}f^2\tilde b \label{eqb}
\ena 

Similarly, we will use the change of variable \fbox{$\ds f(u)= 
\frac{m_0 t(\sqrt{u})}{\sqrt{u}}$} in Eq.~\ref{sqed-gauge-invar1b} 
and in Eq.~\ref{eqb}. We finally obtain this system of coupled 
differential equations: 

\bea
t''(y) - \left(\frac{m^2}{m_0^2}
-e^2\frac{\tilde b(y)^2}{y^2}\right)t 
&=& 0 \label{eqt} \\ 
\tilde b''(y) -2e^2\frac{t(y)^2}{y^2}\tilde b(y) 
&=& 0 \label{eqb2}
\ena

\subsubsection{Normalization of the solutions}

In this paragraph, we compute the conserved momentum   
 of spherically symmetric solitons. We will 
show that the solutions cannot be normalized. Considering 
the energy-momentum tensor of Eq.~\ref{em-tensor2}, we get 
for a soliton: 

\bea
\vec P &=& \vec 0 \\ 
F_{0\lambda}F_0^{~\lambda} 
&=& -4m_0^6 \left(\frac{db}{du}\right)^2 
r^2~~~~(rest~frame,~t=0,~r=|\vec x| ) \\ 
v^0 &=& m_0 b(u)~~~;~~~\partial_0 z =0~~~;~~~
(\partial_\mu z)^2 = -4m_0^2 u \left(\frac{df}{du}\right)^2 \\  
T_{00} &=& e^2z^2(2m_0^2b^2-m_0^2b^2+m_0^2a^2u)
+4m_0^2 u \left(\frac{df}{du}\right)^2 
+m^2 f^2 +4m_0^6\left(\frac{db}{du}\right)^2r^2
-2m_0^4\left(\frac{db}{du}\right)^2u \\ 
&=& m_0^2 e^2 f^2 b^2 + 4m_0^2 u \left(\frac{df}{du}\right)^2 +m^2 f^2
+2m_0^6 \left(\frac{db}{du}\right)^2 r^2 \\ 
&=& e^2m_0^4 \frac{t^2{\tilde b}^2}{y^4} +m^2\frac{t^2}{y^2} 
+4m_0^6 r^2\left(\frac{1}{2y}\frac{d}{dy}\left(\frac{t(y)}{y}
\right) \right)^2 +2m_0^6 r^2\left(\frac{1}{2y}\frac{d}{dy}
\left(\frac{\tilde b(y)}{y} \right) \right)^2 \\ 
&=& e^2m_0^4 \frac{t^2{\tilde b}^2}{y^4} +m^2\frac{t^2}{y^2} 
+m_0^4\left(\frac{d}{dy}\left(\frac{t(y)}{y}
\right) \right)^2 +\frac{m_0^4}{2}\left(\frac{d}{dy}
\left(\frac{\tilde b(y)}{y} \right) \right)^2 \\  
P_0 &=& 4\pi\int_0^\infty r^2\, dr\, T_{00}   \\ 
P_0 &=& \frac{4\pi}{m_0^3}\int_0^\infty y^2\, dy
\left[ e^2m_0^4 \frac{t^2{\tilde b}^2}{y^4} +m^2\frac{t^2}{y^2} 
+m_0^4\left(\frac{d}{dy}\left(\frac{t(y)}{y}
\right) \right)^2 +\frac{m_0^4}{2}\left(\frac{d}{dy}
\left(\frac{\tilde b(y)}{y} \right) \right)^2 \right] 
\nonumber \\ 
&=& 4\pi m_0\int_0^\infty dy\left[ e^2 \frac{t^2{\tilde b}^2}{y^2}
+\left(\frac{m}{m_0}\right)^2 t^2 
+y^2\left(\frac{d}{dy}\left(\frac{t(y)}{y}
\right)\right)^2
 +\frac{y^2}{2}\left(\frac{d}{dy}\left(\frac{\tilde b(y)}{y}
\right)\right)^2 
  \right] \label{solit-norm}
\ena

We have seen that $\tilde b$ must tend to a 
non-vanishing constant 
at infinity in space-like directions ($A^\mu \simeq 1/r$), 
but from Eq.~\ref{eqb2} 
we can conclude that $\tilde b$ is a convex 
function when $\tilde b>0$ and 
the converse for the other sign. From the last term 
in Eq.~\ref{solit-norm}, we get that $\tilde b$ cannot tend 
to a non-vanishing value in $y=0$ (otherwise the integral is
divergent).  Thus if $\tilde b$ vanish 
in $y=0$, it cannot tend to a non-vanishing constant at 
infinity because it is a convex function if $\tilde b>0$ or  
the converse if $\tilde b<0$. The only possibility 
is $\tilde b=0$, and we are then back to the free case, which  
we have previously rejected. 

\subsection{Is there some periodic solutions to the coupled equations?}
\label{periodic}

Now we introduce a ``time'' variable $\tau = p\cdot x$ which 
is dimensionless and $y=\sqrt{u}$ like in the soliton case. We have: 

\bea
z(x) &=& f(\tau,y) = \frac{t(\tau,y)}{y}\\ 
v^\mu &=& a(\tau,y)\lambda^\mu +b(\tau,y)p^\mu \\ 
\Rightarrow \partial_\mu z &=& p_\mu \partial_0 f
+\frac{\lambda^\mu}{y}\partial_1 f \\ 
\Rightarrow \Box z &=& m_0^2\left(\partial_0^2 f -\frac{2}{y}\partial_1 f 
-\partial_1^2 f \right) 
= \frac{m_0^2}{y}\left(\partial_0^2 t -\partial_1^2 t \right) \\ 
\Box v_\mu -\partial_\mu(\partial\cdot v) &=& 
m_0^2 p_\mu \left[\rho\partial_0\partial_1 a +3\partial_0 a 
-\partial_1^2 b -\frac{2}{y}\partial_1 b \right]
+m_0^2 \lambda_\mu\left[\partial_0^2 a 
-\frac{\partial_0\partial_1 b}{y} \right] 
\ena

From these basic calculations we get for the equations of motion: 

\bea
\partial_0^2 t -\partial_1^2 t 
+\left(\frac{m^2}{m_0^2}-e^2(b^2-y^2 a^2)\right)t &=& 0 \label{eq1}\\ 
\partial_0^2 a 
-\frac{\partial_0\partial_1 b}{y} &=& -2\frac{e^2 t^2}{m_0^2 y^2}a 
\label{eq2} \\ 
 y\partial_0\partial_1 a +3\partial_0 a 
-\partial_1^2 b -\frac{2}{y}\partial_1 b &=& 
-2\frac{e^2 t^2}{m_0^2 y^2}b  \label{eq3}
\ena

These equations are much more complicated than in the case of 
 solitons and the fundamental structure of the solutions, 
even periodic in time is not clear so far. We will restrict 
ourselves in this paragraph to a description of what is 
really different in this case and why we conjecture 
the existence of some normalized periodic solutions. \\ 

The conservation of the electromagnetic current leads 
to the emergence of a kind of pre-potential:  

\bea
\partial_\mu(z^2 v^\mu) &=& 0 \\ 
\Leftrightarrow \partial_1 (y t^2 a) &=& \partial_0 (t^2 b) 
\label{conserved-c} \\ 
\Rightarrow y t^2 a &=& \partial_0\vp(\tau,y) \label{vpa} \\ 
and~~~~~~~  t^2 b &=& \partial_1\vp(\tau,y) \label{vpb}
\ena

Introducing this potential in the equations for the 
electromagnetic field we get: 

\bea 
\partial_0\left( \partial_0 a 
-\frac{\partial_1 b}{y} \right) &=& -2\frac{e^2}{m_0^2}
\partial_0\left(\frac{\vp}{y^3}\right)   
\label{eq2b} \\ 
 \partial_1\left(y^2\partial_1 b\right)  
-\partial_1\partial_0 (y^3a) &=& 
2\frac{e^2}{m_0^2}\partial_1\vp \label{eq3b}
\ena

These equations can be partially integrated, and 
we obtain:

\bea 
 \partial_0 a 
-\frac{\partial_1 b}{y} &=& -2\frac{e^2}{m_0^2}
\frac{\vp}{y^3} +A(y)  
\label{eq2c} \\ 
 y^2\partial_1 b -\partial_0 (y^3a) &=& 
2\frac{e^2}{m_0^2}\vp +B(\tau) \label{eq3c}
\ena

The presence of these two functions A and B 
 enlarges significantly the set of possibilities for the 
solutions. We therefore hope that some of 
these might be normalizable, as we shall discuss further. 

\subsubsection{Normalization of the time-dependent solutions}

In order to normalize these periodic solutions, the 
computation of the conserved momentum gives for 
Eq.~\ref{em-tensor2}: 

\bea
F_{\mu\nu}&=& (p\wedge\lambda)_{\mu\nu}\left(\partial_0 a - 
\frac{\partial_1 b}{y}\right) \\ 
F_{\mu\alpha}F_\nu^{~\alpha} &=& -m_0^2\left(\partial_0 a - 
\frac{\partial_1 b}{y}\right)^2(y^2 p_\mu p_\nu
-\lambda_\mu\lambda_\nu) \Rightarrow F_{0\alpha}F_\mu^{~\alpha} 
= -m_0^3y^2\left(\partial_0 a - 
\frac{\partial_1 b}{y}\right)^2 p_\mu \\ 
F_{\beta\alpha}F^{\beta\alpha} &=& -2m_0^4 y^2\left(\partial_0 a - 
\frac{\partial_1 b}{y}\right)^2 \\ 
T_{0i} &=& 2e^2f^2m_0ab\lambda^i 
+2m_0\partial_0f \frac{\lambda^i}{y} \\ 
\Rightarrow P_i &=& 0 \\ 
T_{00} &=& e^2m_0^2f^2(b^2+a^2y^2)+m^2f^2
+m_0^2\left( (\partial_0f)^2+(\partial_1f)^2 \right) 
+\frac{y^2}{2}\left(\partial_0 a - 
\frac{\partial_1 b}{y}\right)^2 \\  
P_0 &=& \frac{4\pi}{m_0^3}\int_0^\infty y^2dy\, T_{00} \\ 
&=& \frac{4\pi}{m_0}\int_0^\infty y^2dy\, 
\left( e^2f^2(b^2+a^2y^2)+\frac{m^2}{m_0^2}f^2
+\left( (\partial_0f)^2+(\partial_1f)^2 \right) 
+\frac{y^2}{2m_0^2}\left(\partial_0 a - 
\frac{\partial_1 b}{y}\right)^2 
 \right) \label{periodic-energy} 
\ena

The last term in  Eq.~\ref{periodic-energy} 
 also appears in Eq.~\ref{eq2c}, equation 
that was absent when we considered soliton solutions. In this 
equation, the function A is 
 undetermined but if $\frac{\vp}{y^3}$ 
is sufficiently singular at 0, the function A will certainly not 
compensate the singularity because it is time-independent, and 
$\vp$ is periodic. Thus, if A accidentally compensate 
$\frac{\vp}{y^3}$ at $y=0$ for $t=0$, it may not be the case at a 
different time. As a consequence, $\vp$ must certainly vanish at 
$y=0$ if one wants the integral to be convergent. 

We still have in this case some dramatic constraints 
on the behaviour of the solutions at $y=0$. However, 
what prevented us from finding normalized periodic solutions 
in the free case was the finite value of t at $y=0$. 
In the free case, solutions are only composed of exponentially 
decreasing functions. Here we have another ``mass'' term 
in the equation of motion for t. If $b^2-a^2y^2$ becomes large 
in the vicinity of the origin, one may obtain solutions 
that are spatially oscillating (and only near $y=0$). 
Such a possibility allows to have a t function that vanishes 
at $y=0$, while still featuring an exponentially 
decreasing behaviour at infinity. 
We expect soon to be able to confirm this conjecture by numerical 
simulations, before we can get more rigorous answers 
to this problem. 

%%%%%%%%%%%%%%%%%%%%%%%%%%%%%%%%%%%%%%%%%%%%%%%%%%%%%%%%%%%%%%%%%%%%
%%%%%%%%%%%%%%%%%%%%%%%%%%%% SECTION 3 %%%%%%%%%%%%%%%%%%%%%%%%%%%%%
%%%%%%%%%%%%%%%%%%%%%%%%%%%%%%%%%%%%%%%%%%%%%%%%%%%%%%%%%%%%%%%%%%%%
\section{The non-abelian case}

\subsection{The standard equations of motion}

 In this case we consider a scalar field $\Phi$ lying in an 
N-dimensional vector space of representation of the 
Lie group ${\cal G}$ (a subgroup of $U(N)$). The results 
presented in this section will not work for all the possible 
gauge groups, 
yet our method is valid for $U(N)$ or $SU(N)$. 
There are very few constraints that may 
be imposed on a generic gauge group. 
Probably the most important one 
is that there must exist a scalar product on the Lie algebra 
which is invariant under an inner automorphism. One can then 
demonstrate that the solvable part (in the Levi decomposition 
of the group) must be abelian. Thus if we also 
constrain the group to be compact, it is relevant to consider 
gauge groups as being a sum of $U(1)$ terms, plus 
any semi-simple part like $SU(N)$. Since the $U(1)$ case 
has been previously solved, we focus here on $SU(N)$ groups. 
Actually, we will see that 
our formalism works if the orbit of any vector $\Phi$ under 
the gauge group is $C^N$, which is the case for $SU(N)$. \\  

We will denote by $i{\cal A}$ the real Lie algebra, such 
that the matrices lying in ${\cal A}$ are hermitian. 
If $\rho$ stands for  
the representation of the Lie algebra, 
we can endow the algebra with the following 
scalar product for the computation of the 
Yang-Mills part of the lagrangian: 
$(A,B)_\rho=\tr[\rho(A)^\dagger \rho(B)]$.  
The scalar product generally used with a semi-simple group 
is the Killing form applied to the field strength tensor.  
Since this scalar product is proportional to any 
scalar product of the form $(A,B)_\rho$ (the coefficient 
being the Dynkin index of $\rho$), we will simply use 
this scalar product $(A,B)=\tr[A^\dagger B]= \tr[AB]$.  
 In the following, we give the 
lagrangian and the corresponding equations of motion, using 
$\Phi$ and $W_\mu$ as variables.

\bea
{\cal L} &=& {\cal L}_0 + {\cal L}_{YM} \nonumber \\ 
{\cal L}_0 &=& (D_\mu\Phi)^\dagger D^\mu\Phi - m^2\Phi^\dagger\Phi 
\nonumber \\ 
&=& \partial_\mu\Phi^\dagger \partial^\mu\Phi 
-ig\Phi^\dagger W_\mu\partial^\mu\Phi 
+ig\partial_\mu\Phi^\dagger W^\mu\Phi +g^2\Phi^\dagger W_\mu W^\mu\Phi
- m^2\Phi^\dagger\Phi \\ 
{\cal L}_{YM} &=& -\frac{1}{4}(F_{\mu\nu},F^{\mu\nu}) = 
-\frac{1}{4}\Big\{\tr\left[(\partial_\mu W_\nu-\partial_\nu W_\mu)
(\partial^\mu W^\nu-\partial^\nu W^\mu)\right]  \nonumber \\  
&& -g^2\tr\left[[W_\mu,W_\nu][W^\mu,W^\nu]\right] 
+2ig\tr\left[(\partial_\mu W_\nu-\partial_\nu W_\mu)
[W^\mu,W^\nu]\right]\Big\} \\
\frac{\partial{\cal L}}{\partial(\partial_\mu W_\nu)}(\Omega_\nu) 
&=& -\,\tr[\Omega_\nu G^{\mu\nu}]  \\ 
\partial_\mu \frac{\partial{\cal L}}{\partial(\partial_\mu W_\nu)}
(\Omega_\nu) - \frac{\partial{\cal L}}{\partial(W_\nu)}
(\Omega_\nu) &=& -\,\tr[\Omega_\nu \partial_\mu G^{\mu\nu}] 
-\left(-ig\Phi^\dagger \Omega_\mu\partial^\mu\Phi 
+ig\partial_\mu\Phi^\dagger \Omega^\mu\Phi +g^2\Phi^\dagger 
\{\Omega_\mu, W^\mu\}\Phi
\right) \nonumber \\ 
&& +ig\,\tr\left[[W_\mu,\Omega_\nu]G^{\mu\nu} 
\right]  \\ 
&=& -\,\tr[\Omega_\nu \partial_\mu G^{\mu\nu}] 
+ig\tr\left[\Omega_\nu\left(D^\nu\Phi \Phi^\dagger 
-\Phi (D^\nu\Phi)^\dagger 
\right)\right] \nonumber \\ 
&& +ig\,\tr\left[\Omega_\nu[G^{\mu\nu},W_\mu] 
\right]  \\ 
\Rightarrow 0&=& \tr\left[\Omega_\nu \left(-{\cal D}_\mu G^{\mu\nu} +
ig(D^\nu\Phi\Phi^\dagger -\Phi (D^\nu\Phi)^\dagger) \right) 
\right]~~~~~(\forall~{\Omega_\nu}~\in~{\cal A}) \label{gf-na} \\ 
{\cal D}^\mu (G_{\mu\nu})  
&=&  \Pi_{\cal A}\left[ig\left(D_\nu\Phi\Phi^\dagger 
-\Phi(D_\nu\Phi)^\dagger \right)\right]= \Pi_{\cal A}(J_\nu)
\label{gf-na1} \\ 
J_\mu &=& ig\left(\partial_\mu \Phi \Phi^\dagger -\Phi\partial_\mu
\Phi^\dagger +ig\{ W_\mu, \Phi\Phi^\dagger\}\right) \label{curr} \\ 
\partial_\mu \frac{\partial{\cal L}}{\partial(\partial_\mu \Phi^\dagger)}
 - \frac{\partial{\cal L}}{\partial(\Phi^\dagger)}
=0\Rightarrow 0 &=& (D_\mu D^\mu +m^2)\Phi \nonumber \\ 
&=& (\Box+m^2)\Phi + 2igW_\alpha\partial^\alpha\Phi 
+ig(\partial_\alpha W^\alpha)\Phi - g^2W_\mu W^\mu\Phi 
 \label{scalar-na} 
\ena

 We shall note that in Eq.~(\ref{gf-na}), the 
fact that the equation is only valid for $\Omega_\nu~\in~{\cal A}$ is very 
important. It comes from the fact that in the variationnal 
principle leading to the Euler-Lagrange equations, the variation 
of the gauge field ($\Omega_\nu$) must lie in the Lie algebra also.  
If this equation were valid for any matrix $\Omega_\nu$, then 
we would have ${\cal D}^\alpha (G_{\alpha\beta})$ (which is 
in ${\cal A}$) equal to $ig\left(D_\beta\Phi\Phi^\dagger 
- \Phi(D_\beta\Phi)^\dagger \right)$, which is not necessarily 
in ${\cal A}$, and this is why there is this projection 
operator on the Lie algebra $\Pi_{\cal A}$ in Eq.~\ref{gf-na1}. 
 For $su(N)$ algebras, this 
projection is simply $M\mapsto M -\tr(M)\frac{I}{N}$, where 
$I$ stands for the identity matrix.  
Eq.~\ref{gf-na1} and Eq.~\ref{scalar-na} are the equations of motion 
respectively for the gauge fields and for the scalar fields, which 
can be related to the abelian equations of Eq.~\ref{gf} 
and Eq.~\ref{scalar}. The non-abelian equivalent of the 
current is now extracted from Eq.~\ref{gf-na1} and is 
given by the matrix: $J_\nu = ig\left(D_\nu\Phi\Phi^\dagger 
-\Phi(D_\nu\Phi)^\dagger \right)$ \\ 
Contrary to the abelian case, this current is 
not gauge invariant anymaore but rather gauge covariant, that is 
$J_\nu =U{J'}_\nu U^{-1}$ under a gauge transformation. \\ 

We now operate as in the previous section, and 
observe that if we compute $(\ref{scalar-na})\Phi^\dagger - 
\Phi(\ref{scalar-na})^\dagger$, we get:

\bea
0 &=& (D_\mu D^\mu\Phi) \Phi^\dagger-\Phi(D_\mu D^\mu\Phi)^\dagger 
\Phi \nonumber \\ 
&=& {\cal D}_\mu \left( (D^\mu\Phi) \Phi^\dagger
-\Phi(D^\mu\Phi)^\dagger \right) \\  
\Rightarrow  0 &=& {\cal D}_\mu \left( J^\mu \right) \label{non-ab-cons-c} 
\ena

If one projects this equation on the Lie algebra, the 
resulting equation is redundant with Eq.~\ref{gf-na1} on which 
we apply the operator ${\cal D}^\nu$. Like in the abelian case, 
we find a redundancy, but it is important to note at this stage 
that eq.~\ref{non-ab-cons-c} is a stronger condition than if 
we just applied ${\cal D}^\nu$ on Eq.~\ref{gf-na1}. It seems 
that we missed some degrees of freedom in Eq.~\ref{gf-na1}. 
The fundamental 
structure of the gauge group is responsible for this fact. 
For instance, in the case of a $u(N)$ algebra, 
$\Pi_{\cal A}(M)=M$ if 
M is hermitian, and all the ``degrees of freedom'' of $J_\mu$ 
are concerned with this redundancy between the equation 
for the matter and the equation for the gauge field.  \\  

We therefore have too much information in the set of equations 
of the matter field and one should replace eq.~(\ref{scalar-na}) by 
$(\ref{scalar-na})\Phi^\dagger +  
\Phi(\ref{scalar-na})^\dagger$, i.e.:

\bea
0 &=& (D_\mu D^\mu\Phi)\Phi^\dagger +
\Phi(D_\mu D^\mu\Phi)^\dagger  +2m^2 \Phi\Phi^\dagger  \\
&=& {\cal D}_\mu \left((D^\mu\Phi) \Phi^\dagger
+\Phi(D^\mu\Phi)^\dagger \right) 
-2D_\mu\Phi(D^\mu\Phi)^\dagger+2m^2\Phi\Phi^\dagger\\ 
&=& \left({\cal D}_\mu{\cal D}^\mu(\Phi\Phi^\dagger) 
-2(D_\mu\Phi)(D^\mu\Phi)^\dagger+2m^2\Phi\Phi^\dagger\right) 
\ena
   
\subsection{The gauge invariant variables}

The procedure used to obtain Eq.~\ref{ym-from-c} consists in 
eliminating the two first terms of $\frac{J_\mu}{ie} = 
\vp^*\partial_\mu\vp-\partial_\mu\vp^*\vp +2ie\rho A_\mu$ in order 
to extract the gauge field. We have $\frac{J_\mu}{ie\rho}=2ie 
A_\mu +\partial_\mu \Lambda$ and the pure gauge term disappears in 
$F_{\mu\nu}$. But in our case we have a matrix and this procedure 
does not work. However, the extraction of $A_\mu$ can be seen in 
another way. In the abelian case, we could also have taken a 
unitary gauge, that is to say a gauge in which $\vp$ is real. This 
automatically eliminates the desired terms. We may proceed here in 
a similar way. The essential hypothesis is that any two scalar 
fields $\Phi$ and $\Psi_0$ can be related by an element of the 
gauge group. It is the case for $U(N)$ or $SU(N)$. Thus, the 
central point of the method is to choose a constant unitary vector 
$\Psi_0$, and therefore one can find $U$ in the gauge group such 
that: 

\bea
\Phi &=& zU\Psi_0 \\ 
z &=& \sqrt{\Phi^\dagger\Phi} = \sqrt{\rho} 
\ena 
 
A consequence is that if ${W'}_\mu$ is the gauge field in the 
``unitary'' gauge obtained by the matrix $U$ we have 
from Eq.~\ref{curr}: 

\bea
{J'}_\mu = U^{-1}J_\mu U &=& ig\left( z\partial_\mu(z) 
\Psi_0\Psi_0^\dagger - \Psi_0\Psi_0^\dagger z\partial_\mu(z) 
+ig\rho\{{W'}_\mu,\Psi_0\Psi_0^\dagger\} \right)\\ 
&=& (ig)^2\rho \{{W'}_\mu,\Psi_0\Psi_0^\dagger\} 
\ena

However, it is in general impossible to reconstruct the 
entire gauge field ${W'}_\mu$ from this equation, 
except for the $SU(2)$ case because of the relation 
$\{\sigma^i,\sigma^j\}=2\delta^{ij}$ (and this 
anticommutator has no residue 
lying in the Lie algebra). Using this property, the traceless 
part of ${J'}_\mu$ gives $(ig)^2\rho {W'}_\mu$. Since it works 
only for $SU(2)$, we need to find a way to get the missing 
degrees of freedom of the gauge field. The method consists 
in constructing an orthonormal basis of $C^N$, starting 
from $\Psi_0$: $(\Psi_0,\Psi_1,...,\Psi_{N-1})$, which does not 
depend on space-time coordinates. If we set $\Phi_k= zU\Psi_k$ 
($k\geq 1$), then $(\rho^{-1}\Phi,\rho^{-1}\Phi_1,...,\rho^{-1}\Phi_{N-1})$ 
forms also an orthonormal basis. A gauge transformation will 
naturally apply also to these new scalar fields, and 
we consider the gauge invariant variables: 

\bea
J_{mn\mu} &=& ig\left(\Phi_m^\dagger D_\mu\Phi_n 
- (D_\mu\Phi_m)^\dagger\Phi_n\right) 
\ena

The simple reason why we do not consider some other 
gauge invariant variables, by taking the sum of the two terms 
above instead of their difference is that 
$\Phi_m^\dagger D_\mu\Phi_n + (D_\mu\Phi_m)^\dagger\Phi_n= \partial_\mu 
(\Phi_m^\dagger\Phi_n) =\partial_\mu(\rho\delta_{m,n})$ 
and thus they can be expressed using the gauge invariant  
variable $z=\sqrt{\rho}$. In the unitary gauge, these gauge 
invariant variables allow to reconstruct the 
gauge field completely: 

\bea
J_{mn\mu} &=& 2(ig)^2\rho\Psi_m^\dagger{W'}_\mu\Psi_n 
\quad\quad\quad (J_{nm\mu} = J_{mn\mu}^*) \\ 
v_{mn\mu} &=& \frac{-1}{2g^2\rho}J_{mn\mu} \\ 
\Rightarrow {W'}_\mu &=& \sum_{m,n} v_{mn\mu}\Psi_n\Psi_m^\dagger  
\ena

The equations of motion for the gauge field in Eq.~\ref{gf-na1} can 
then be rewritten in the unitary gauge (note that 
$\Pi_{\cal A}(U^{-1}JU)= U^{-1}\Pi_{\cal A}(J)U$) 

\bea
{G'}_{\mu\nu} &=& \sum_{m,n} \left(
\partial_\mu\left(v_{mn\nu}\right) - 
\partial_\nu\left(v_{mn\mu}\right)
+ig \sum_k (v_{mk\mu}v_{kn\nu} - v_{mk\nu}v_{kn\mu}) 
\right)\Psi_n\Psi_m^\dagger \label{curvature} \\ 
{\cal D'}^\mu({G'}_{\mu\nu}) &=& \partial^\mu{G'}_{\mu\nu} 
+ig[{W'}_\mu,{G'}_{\mu\nu}] \\ 
&=&\sum_{m,n} \left(
\Box \left(v_{mn\nu}\right) - 
\partial_\nu\left(\partial\cdot v_{mn}\right)
+ig \sum_k \partial^\mu (v_{mk\mu}v_{kn\nu} - v_{mk\nu}v_{kn\mu}) 
\right)\Psi_n\Psi_m^\dagger \nonumber \\ 
&&+ig\sum_{m,n} \Psi_n\Psi_m^\dagger \left[\sum_l  v_{ml}^{~~\mu}
 \left(
\partial_\mu\left(v_{ln\nu}\right) - 
\partial_\nu\left(v_{ln\mu}\right)
+ig \sum_k (v_{lk\mu}v_{kn\nu} - v_{lk\nu}v_{kn\mu}) 
\right)\right. \nonumber \\ 
&&\left. -\left(
\partial_\mu\left(v_{ml\nu}\right) - 
\partial_\nu\left(v_{ml\mu}\right)
+ig \sum_k (v_{mk\mu}v_{kl\nu} - v_{mk\nu}v_{kl\mu}) 
\right)
v_{ln}^{~~\mu}\right] \\ 
&=& -g^2z^2 \sum_m \Pi_{\cal A}\left(v_{0m\mu}
\Psi_m\Psi_0^\dagger +v_{m0\mu}\Psi_0\Psi_m^\dagger 
\right) \\ 
&=& -g^2z^2 \sum_m \left(v_{0m\mu}
\Psi_m\Psi_0^\dagger +v_{m0\mu}\Psi_0\Psi_m^\dagger 
 -2\frac{v_{00\mu}}{N}\Psi_m\Psi_m^\dagger\right) 
\ena

The last equality is only valid for $SU(N)$. One must adapt this 
formula for another gauge group. Projecting these equations on the 
basis of matrices $\Psi_m\Psi_n^\dagger$ leads to a large set of 
$N^2$ equations in which only gauge invariant variables are 
present. In the $SU(N)$ case we can also separate these 
equations into four different classes depending on the indices m 
and n, because of the specific form of the current matrix 
projected on the Lie algebra. The four cases correspond to  the 
diagonal case with indices in the form $(m,m)$ ($m>0$), the case 
with indices in the form  $(0,m)$ or $(m,0)$ ($m>0$), and finally 
the case where $m=n=0$. The projection on these different cases 
can be easily done and we will not present them here. It is clear 
that the gauge fields ${W'}_\mu$ expressed in the basis of the 
$\Psi_k$'s is nothing but the matrix composed of the gauge 
invariant coefficients $(v_{m,n\mu})$. Of course, these 
coefficients depend on the constant basis we choose, but physical 
solutions must be independent of this choice. 
It remains to demonstrate that these equations of motion can be 
re-expressed using only variables that are also independent from 
the constant basis chosen: we can consider some objects of the 
form $\tr[({W'}_\mu)^n]$, or equivalently the characteristic 
polynomial of ${W'}_\mu$.  We expect to have new results in the 
near future. 

We may conclude this 
last section with the equation of motion for the 
matter fields. The simplest way is to look at the lagrangian 
and to use the following equality: 

\be
\frac{1}{-g^2}\sum_{m,n}J_{mn\mu}J_{nm}^{~~\mu} = N\partial_\mu \rho 
\partial^\mu \rho -4\rho (D_\mu \Phi)^\dagger (D^\mu\Phi) 
\ee   

The matter part of the lagrangian can then be written: 

\be
{\cal L}_0 = N\partial_\mu z \partial^\mu z -m^2z^2  
+g^2z^2\sum_{m,n}v_{mn\mu}v_{nm}^{~~\mu} 
\ee

And the equation of motion for the scalar field is finally: 

\be
\left(\Box+\frac{m^2}{N}\right)z = 
\frac{g^2}{N}z \sum_{m,n}v_{mn\mu}v_{nm}^{~~\mu}  
\ee

%%%%%%%%%%%%%%%%%%%%%%%%%%%%%%%%%%%%%%%%%%%%%%%%%%%%%%%%%%%%%%%%%%%%%%
%%%%%%%%%%%%%%%%%%%%%%%%%%%% CONCLUSION %%%%%%%%%%%%%%%%%%%%%%%%%%%%%%
%%%%%%%%%%%%%%%%%%%%%%%%%%%%%%%%%%%%%%%%%%%%%%%%%%%%%%%%%%%%%%%%%%%%%%
\section{Conclusion}

In this paper, we give a certain number of results which 
are really encouraging 
for the purpose of reformulating gauge theories using only 
gauge-invariant variables. 
Within  the prospects of this work, a short-term project would 
naturally be to find an equivalent formulation when fermions are 
involved. Then, the quantization of the theory has to be 
constructed. Within this subtopic, it would be interesting to 
revisit the general formalism of quantization in QFT. An equation 
like $\frac{dA}{dt}= i[H,A]$ is a very old non-relativistic 
formula which is surprisingly still used in textbooks about 
relativistic quantum field theory. Instead of the Hamiltonian, one 
would naturally consider an operator of the form $\int_\Sigma 
d\sigma^\mu T_{\mu\nu}$ in order to quantize a theory. This has 
not been done yet and one of the possible reasons is that there is no 
unique expression for the energy-momentum tensor $T_{\mu\nu}$.  
There are some current research activities on this 
topic~\cite{gotay1}, in order to find the ``best'' criteria to 
define uniquely $T_{\mu\nu}$. So far, it seems that the Belinfante 
tensor is a good candidate, since it is gauge-invariant. Therefore 
it can be naturally inserted in the formalism presented in this 
paper. Finally, in the long-term we hope 
to be able to compute some scattering 
cross sections using directly gauge invariant variables, and 
also to provide a revised version of Quantum Field Theory which 
would apply to unstable particles and more generally, to physical 
systems that evolve  on a ``long-time'' scale, 
(CP violation, neutrino oscillations,...) as mentionned in the 
introduction.  \\ 

%%%%%%%%%%%%%%%%%%%%%%%%%%%%%%%%%%%%%%%%%%%%%%%%%%%%%%%%%%%%%%%%%%%%%%
%%%%%%%%%%%%%%%%%%%%%%%%%%%% Appendices %%%%%%%%%%%%%%%%%%%%%%%%%%%%%%
%%%%%%%%%%%%%%%%%%%%%%%%%%%%%%%%%%%%%%%%%%%%%%%%%%%%%%%%%%%%%%%%%%%%%%
\section{Appendices}

\subsection{Review of basic cohomological formulas}

As noted in this paper, one of the main problems regarding 
 gauge independence is to have  a method to find the set of 
gauge fields with a given Field-Strength tensor $F^{\mu\nu}$.  
We will separate the abelian case 
from the non-abelian one, because the curvature tensor 
depends linearly on the gauge field in the abelian case, 
quadratically in the latter 
case. Linearity is lost in the non-abelian case, which renders 
the problem much more complicated.   \\ 
 
The problem can be summarized as follows: if one has a specific 
tensor F of rank $n$, we look for another tensor of rank $n-1$ 
such that $F=dA$ where $d$ represents the 
exterior derivative. 
The tensor $F$ must obey $dF=0$ because of the property $d^2=0$. 
So we want to find $A$ from a given $F$, 
assumed that $F$ is a closed form (i.e. 
$dF=0$). Given a solution $A$, one can find another solution 
$A'$ by adding to $A$ any term of the form $d\Lambda$, again 
because $d^2=0$. Therefore, we will say that two tensors 
of rank $n-1$ are co-homologous if there exists $\Lambda$ such 
that $A-A' = d\Lambda$. It is an equivalence relation 
and the equivalence classes are called 
cohomology classes 
(for the de-Rahm cohomology, and we will 
further explain why it is important to make this 
distinction when the non-abelian case is involved). 
 
\subsection{Abelian gauge fields}

Let $M=R^4$ be the Minkowsky space-time, and consider 
$X^\mu(u,x)$ an application from $[0,1]\times M$ into $M$ 
such that:

\bea
\forall~x~\in~M,~X^\mu(0,x) &=& x_0^\mu \label{contraction00a} \\ 
\forall~x~\in~M,~X^\mu(1,x) &=& x^\mu \label{contraction01a}
\ena

 We also assume that $X^\mu$ is infinitely smooth. It is 
then called a ``contraction''. The reader will recover the 
standard Poincar\'e formula by taking $X^\mu(u,x)=ux^\mu$.  
Suppose $A^\mu(x)$ is a vector field with vanishing curvature, 
then if we define $V(x)$ as follows: 

\bea 
\label{de-rahm0}
 V(x) &=&  \int_0^1 du \frac{\partial X^\mu}{\partial u} 
A_\mu(X(u,x)) \\ 
Then~~~~~~~~~~~\partial_\mu V &=&  A_\mu(x) -\int_0^1 du
\frac{\partial X^\alpha}{\partial u}\frac{\partial X^\beta}
{\partial x^\mu}F_{\alpha\beta}(X)\label{de-rahm00}
\ena 

 Therefore, if the curvature 
of $A$ vanishes, $V(x)$ is a possible solution for the potential. 
Also, if one replaces explicitly $A^\mu$ by $\partial^\mu V'$ in 
eq.~\ref{de-rahm0}, one gets $V'(x)-V'(x_0)$, and not $V'(x)$. 
$V(x)$ is therefore not a ``fixed point solution'' of an integral 
equation, but can be defined as the solution for which 
$V(x_0) = 0$. The rest in the expression of $\partial_\mu V$ 
vanishes explicitly for a vanishing curvature, but when the 
curvature is not $0$, this formula provides us with an explicit 
expression for $A^\mu$ as a function of $F^{\mu\nu}$ up to a gauge 
transformation by $\partial^\mu V$. Thus we have already the 
next step, and if we consider a given field-strength tensor 
$F^{\mu\nu}$, we can define the following vector field:

\be
\label{de-rahm1}
A^\mu(x) = \int_0^1 du \frac{\partial X^\alpha}{\partial u} 
\frac{\partial X^\beta}{\partial x^\mu} F_{\alpha\beta}(X(u,x)) 
\ee

Then, with this definition we have:

\be
\partial_\mu A_\nu - \partial_\nu A_\mu = F_{\mu\nu}(x) 
-\int_0^1 du \frac{\partial X^\alpha}{\partial u} 
\frac{\partial X^\beta}{\partial x^\mu}\frac{\partial X^\gamma}
{\partial x^\nu}(\partial_\alpha F_{\beta\gamma} + 
 \partial_\beta F_{\gamma\alpha} + \partial_\gamma F_{\alpha\beta})  
\ee

The last term vanishes if $dF=0$, and we recognize here the 
homogeneous Maxwell equations. 
In this case, the expression we have 
chosen for $A^\mu$ is a possible gauge field, and this 
formula is of course very important because it allows us 
to ``parameterize'' the orbits of gauge fields.
 It is possible to go on with this scheme, and for a given 
3-form $\omega_{\alpha\beta\gamma}$ we can define $F^{\mu\nu}$ 
using:

\be
F_{\mu\nu} = \int_0^1 \frac{\partial X^\alpha}{\partial u} 
\frac{\partial X^\beta}{\partial x^\mu}\frac{\partial X^\gamma}
{\partial x^\nu}\omega_{\alpha\beta\gamma}(X(u,x))
\ee

and when $d\omega =0$, we have $\partial_\alpha F_{\beta\gamma} + 
 \partial_\beta F_{\gamma\alpha} + \partial_\gamma F_{\alpha\beta}
= \omega_{\alpha\beta\gamma}(x)$, and so on (but there is actually 
only one next step because we have assumed here that we are in 
four space-time dimensions and any four form is proportional 
to the Levi-Civita pseudo-tensor\index{Levi-Civita pseudo-tensor}). 

To summarize, given an $n$ form $F$ such that $dF=0$, we have been 
able to exhibit a $n-1$ form $A$ such that $F=dA$. This element 
$A$ can be interpreted as an element of an equivalent class of 
cohomology with a given curvature. In other words, we have 
``computed'' the cohomology. Expressed this way, it looks simple 
but hides the real difficulties, which are of a topological nature. 
In all these calculations, we have assumed the existence of $X^\mu$, 
which imposes some constraints on the topology of the four dimensional 
space-time. If the whole Minkowsky space is 
taken under consideration, no topological problem occurs, 
and more generally, this is true if we consider a 
simply connected space. Then, one 
can find $X^\mu$ and proceed to the previous calculations. 

\subsection{Conventions for the Non-abelian case}

$i{A}$ and $iB$ are supposed to lie in the real Lie 
algebra corresponding 
to the Lie Group\index{Lie Group} ${\cal G}$, 
which is a subgroup\index{Subgroup} of $U(N)$ here. 
Therefore $A$ and $B$ are hermitian. We set ${A}= U{A}'U^{-1}$. 
$X$ and $Y$ are vectors lying in the same representation 
as the matter field $\Phi$.  

\bea
\Phi &=& U\Phi' = e^{iT}\Phi'~~~(T~~small)\label{gt01} \\
W_\mu &=& U{W'}_\mu U^{-1} +\frac{i}{g}\partial_\mu (U) U^{-1}
~~~~~~{W'}_\mu = U^{-1}W_\mu U-\frac{i}{g} U^{-1}\partial_\mu (U) 
\label{gt02} \\ 
\delta W_\mu = {W'}_\mu-W_\mu  &=& -\frac{i}{g}{\cal D'}_\mu(U) U^{-1} 
 = -\frac{i}{g}U^{-1}{\cal D}_\mu(U) \label{gt03} \\ 
U{W'}_\nu U^{-1}-{W'}_\nu +\frac{i}{g}(\partial_\nu U)U^{-1} 
&=& e^{iT}{W'}_\nu e^{-iT}-{W'}_\nu+\frac{i}{g}(\partial_\nu e^{iT})
e^{-iT} \\ 
&\simeq & [iT,{W'}_\nu] -\frac{1}{g}(\partial_\nu T) = -\frac{1}{g}
{\cal D}_\nu ({A}) \label{gt03b}\\ 
D_\mu \Phi &=& (\partial_\mu +igW_\mu)\Phi \Rightarrow  
D_\mu \Phi = D_\mu (U\Phi') = U {D'}_\mu \Phi' \label{gt04} \\
{\cal D}_\mu ({A}) &=& \partial_\mu{A} +ig[W_\mu,{A}] 
\Rightarrow {\cal D}_\mu (U{A}'U^{-1}) = 
U ({\cal D'}_\mu {A}')U^{-1} \label{gt05} \\ 
{\cal D}_\mu (AB) &=& {\cal D}_\mu (A)B + A{\cal D}_\mu (B) \\ 
{\cal D}_\mu (XY^\dagger) &=& {D}_\mu (X)Y^\dagger 
+ X({D}_\mu (Y))^\dagger \label{gt08}\\
D_\mu (AX) &=& {\cal D}_\mu (A)X + A{D}_\mu (X)\label{gt09} \\ 
{[D_\mu,D_\nu]}\Phi &=& ig(\partial_\mu W_\nu - \partial_\nu W_\mu 
+ig[W_\mu,W_\nu])\Phi = igG_{\mu\nu}\Phi \label{gt06} \\
G_{\mu\nu} &=& U{G'}_{\mu\nu}U^{-1} \label{gt07} \\ 
\ [{\cal D}_\alpha,{\cal D}_\beta](A) 
&=& ig[G_{\alpha\beta},A] \label{gt14} \\  
0 &=& [D_\nu,[D_\rho,D_\sigma]]\Phi+[D_\rho,[D_\sigma,D_\nu]]\Phi+ 
[D_\sigma,[D_\nu,D_\rho]]\Phi  \label{gt15} \\ 
\Leftrightarrow 0 &=& \eps^{\mu\nu\rho\sigma}[D_\nu,G_{\rho\sigma}](\Phi) 
~~~~~~~(\forall\, \Phi) \\ 
\Leftrightarrow 0 &=& {\cal D}_\nu 
(\tilde G^{\mu\nu})~~~~~(Bianchi) \label{gt16}   
\ena

For $SU(N)$ gauge groups, it may be useful to use the relation: 

\be
\Phi\Phi^\dagger = \Phi^\dagger\Phi\frac{1}{N} I + A_\phi 
\label{alg-decomp} 
\ee

where $I$ stands for the identity matrix in $N$ dimensions, 
$A_\Phi$ lies therefore in the Lie algebra
$su(N)$, and we will conveniently denote by 
$\rho_\Phi= \Phi^\dagger\Phi$ the probability density 
of $\Phi$. \\

\subsection{Non abelian case and the Path Ordered Exponential} 

If $A$ is an operator valued function of the real variable 
$\lambda$, a solution to the differential equation 
$f'(\lambda) = A(\lambda)f(\lambda)$ is given by (see~\cite{karp}): 

\bea
f(x) &=& \left[1+\int_0^x d\lambda\, A(\lambda)  
+\int_0^x d\lambda_1\, A(\lambda_1)
 \int_0^{\lambda_1} d\lambda_2\, A(\lambda_2)
+\ldots \right. \nonumber \\  
&&\left. + \int_0^x d\lambda_1\, A(\lambda_1)  
\cdot\ldots\cdot 
\int_0^{\lambda_{n-1}} d\lambda_n\, A(\lambda_n) \right]f(0) \\ 
&=&  \expf^{\int_0^x d\lambda\, A(\lambda)}f(0) =  
 \lim_{ds\rightarrow 0}\prod_{k=n}^1 e^{A(s_k)ds}f(0)\quad 
\quad with \quad s_k=\frac{k\times x}{n}\label{eproduct} \\ 
\expf^{\int_0^x d\lambda\, A(\lambda)} &=& 
\expf^{\int_y^x d\lambda\, A(\lambda)}
\expf^{\int_0^y d\lambda\, A(\lambda)} \\   
\frac{d}{ds} \expf^{\int_0^s F(v)dv} &=& 
F(s) \expf^{\int_0^s F(v)dv} \quad \quad 
\frac{d}{ds} \expf^{\int_s^1 F(v)dv} =  
-\expf^{\int_s^1 F(v)dv}F(s) \label{expfprop01}
\ena 

Note that the product in Eq.~\ref{eproduct} 
is done ``from right to left''. In the following, we list  
a few properties of the path order exponential:  

\bea
\left(\expf^{\int A}\right)^{-1} = \expb^{-\int A} 
&=& 1-\int_0^1 A(u)du +\int_0^1 du_1\,\int_0^{u_1} du_2 A(u_2)A(u_1) 
+\ldots \\ 
\expf^{\int_a^x A'(s)A^{-1}(s)ds} &=& A(x)A^{-1}(a) \label{o-exp-t01a} \\ 
P(x) = \expf^{\int_a^x A(s)ds}\Rightarrow 
\expf^{\int_a^x A(s)+B(s)ds} &=& 
P(x)\expf^{\int_a^x P^{-1}(s)B(s)P(s)ds} \label{o-exp-t02a} \\ 
A(x) \expf^{\int_a^x B(s)ds}A^{-1}(a) &=& 
\expf^{\int_a^x \left(A'(s)A^{-1}(s)+A(s)B(s)A^{-1}(s)\right)ds} 
\label{o-exp-t03a} \\ 
\frac{\partial}{\partial\lambda}\expf^{\int_a^b A(u,\lambda)du} 
&=&  \int_a^b ds\, \expf^{\int_s^b A(u,\lambda)du}\, 
\frac{\partial A(s,\lambda)}{\partial\lambda}
\,\expf^{\int_a^s A(u,\lambda)du}\label{o-exp-t04a}  
\ena

The last formula can be demonstrated easily if one uses 
the product form of the ordered exponential (Eq.~\ref{eproduct})

\subsubsection{Introduction of a space-time contraction}
  
If we now consider a  contraction $X_\mu(u,x)$ where 
$X_\mu(0,x)=x_0$ and $X_\mu(1,x)=x$ (see Eq.~\ref{contraction01a}), 
we obtain the following definition: 

\bea
F(u,x) &=& \frac{\partial X_\mu}{\partial u}(u,x)
A^\mu(X_\mu(u,x)) \label{Fdef} \\ 
\partial_\mu X_\alpha A^\alpha(X_\mu(u,x))|_{u=1}&=& A_\mu(x) 
\label{Fdefb} \\ 
f(x)= \expf^{ig\int_\gamma dl^\mu A_\mu}(x) &=& 
1+(ig)\int_0^1 du\, F(u,x)  
+(ig)^2\int_0^1 du_1\, F(u_1,x)
\int_0^{u_1} du_2 F(u_2,x)
+\ldots \nonumber \\  
&& + (ig)^n\int_0^1 du_1\, F(u_1,x)  
\cdot\ldots\cdot 
\int_0^{u_{n-1}} du_n\, F(u_n,x) 
+\ldots \label{ordered-exp02a} \\ 
&=& \sum_k  (ig)^k\int_{[0;1]^k} du_1...du_k \theta(u_1,\ldots,u_k) 
F(u_1,x)\ldots F(u_k,x) \label{ordered-exp02b} \\ 
&=& \exp\left( ig\int_0^1du\, F(u,x)\right)~~~~(if~[A(x),A(x')]=0) 
\label{ordered-exp02c} \\ 
 \theta(u_1,\ldots,u_k) &=& 1~~~iff~~~~u_1\geq u_2 \ldots \geq u_k, 
~~0~~if~~not \label{ordered-exp02d} \\   
&=& H(u_1-u_2)H(u_2-u_3)...H(u_{k-1}-u_k) \label{ordered-exp02e}
\ena

Each term in the sum can be obtain by the following recursion:

\bea 
J_0(a,b,x)&=&1 \label{recurr01a}\\ 
J_n(a,b,x)&=&\int_a^b ds\, 
F(s,x)J_{n-1}(a,s)  \label{recurr02a} \\ 
J_n(a,a,x)&=& 0 ~~~~\forall~n,x \label{recurr02b} \\
J_n(a,b,x) &=&\int_a^b ds\, \partial_s X_\mu
A^\mu(X_\mu(s,x)) J_{n-1}(a,s)  \label{recurr02c} \\ 
\ena

Let $\Phi$ be a solution (if it exists) to the system of PDE  
$\partial_\mu\Phi=-igW_\mu(x)\Phi$, then:  

\bea
\frac{\partial X^\mu}{\partial u}\partial_\mu\Phi&=&-ig
\frac{\partial X^\mu}{\partial u}W_\mu(x)\Phi \\ 
\frac{d}{du} \Phi(X(u,x)) &=& F(u,x)\Phi(X(u,x)) \\ 
\Rightarrow \Phi(X(u,x)) &=& \expf^{\int_0^u dv\,F(v,x)}\Phi_0 \\ 
\Rightarrow \Phi(x) =\Phi(X(1,x)) &=& 
\expf^{\int_0^1 dv\,F(v,x)}\Phi_0 \\ 
&=& 
\expf^{-ig\int_0^1 du\,\frac{\partial X^\mu}
{\partial u}W_\mu(X(u,x))}\Phi_0 \label{covar-int} 
\ena

If $\Phi$ is a square matrix and $\Phi_0=I$, then $\Phi$ 
is invertible because $\det(\Phi)=e^{-ig\int \tr F}\neq 0$, thus 
$W_\mu = \frac{i}{g}\partial_\mu\Phi \Phi^{-1}$ which is 
a right invariant form, the curvature of which vanishes. It is not 
surprising to get such a constraint. Already in the 
abelian case, if $\phi = e^{-ig\int A}$ then 
$\partial_\mu \phi = -ig\left(A_\mu+\int \partial_u X^\alpha
\partial_\mu X^\beta F_{\alpha\beta}\right)\phi$ (see Eq.~\ref{de-rahm00}) 
and we explicitly show the presence of a curvature term 
as an obstacle to solve the system of differential equations. 
To obtain  a similar formula in the non-abelian case, let us take 
the partial derivatives of Eq.~\ref{covar-int}. We get: 

\bea
\frac{i}{g}\partial_\mu \Phi &=& \int_0^1 ds\, \expf^{-ig\int_s^1 W_x} 
\partial_\mu\left(\frac{\partial X^\nu}
{\partial s}W_\nu(X(s,x))\right) \expf^{-ig\int_0^s W_x} \\ 
&=& \int_0^1 ds\, \expf^{-ig\int_s^1 W_x} 
\left(\partial_\mu\partial_s X^\nu W_\nu(X(s,x)) 
+\partial_s X^\nu \partial_\mu X^\rho 
\partial_\rho W_\nu(X(s,x)) 
\right) \expf^{-ig\int_0^s W_x} \\ 
&=& \int_0^1 ds\, \expf^{-ig\int_s^1 W_x} 
\{\partial_\mu\partial_s X^\nu W_\nu(X) 
+\partial_u X^\nu \partial_\mu X^\rho G_{\rho\nu}(X) \nonumber \\
&& +\partial_s X^\nu \partial_\mu X^\rho 
(\partial_\nu W_\rho(X)-ig[W_\rho,W_\nu])  
\} \expf^{-ig\int_0^s W_x} \\ 
&=& \int_0^1 ds\, \expf^{-ig\int_s^1 W_x} 
\partial_s \left(\partial_\mu X^\nu W_\nu(X)\right) 
\expf^{-ig\int_0^s W_x} \nonumber \\ 
&&+\int_0^1 ds\, \expf^{-ig\int_s^1 W_x} \{
-ig\partial_s X^\nu \partial_\mu X^\rho[W_\rho,W_\nu]  
+\partial_s X^\nu \partial_\mu X^\rho G_{\rho\nu}  
\} \expf^{-ig\int_0^s W_x} \\ 
&=& \int_0^1 ds\,\partial_s\left( \expf^{-ig\int_s^1 W_x} 
 \partial_\mu X^\nu W_\nu(X)  
\expf^{-ig\int_0^s W_x} \right) \nonumber  \\ 
&& -\int_0^1 ds\, \expf^{-ig\int_s^1 W_x}(ig \partial_s
 X^\nu W_\nu(X))\partial_\mu X^\rho W_\rho(X)  
\expf^{-ig\int_0^s W_x} \label{residue01} \\ 
&& +\int_0^1 ds\, \expf^{-ig\int_s^1 W_x}\partial_\mu X^\rho W_\rho(X)  
(ig\partial_s
 X^\nu W_\nu(X))\expf^{-ig\int_0^s W_x} 
\label{residue02} \\ 
&&+\int_0^1 ds\, \expf^{-ig\int_s^1 W_x} \{
-ig\partial_s X^\nu \partial_\mu X^\rho[W_\rho,W_\nu]  
+\partial_s X^\nu \partial_\mu X^\rho G_{\rho\nu}  
\} \expf^{-ig\int_0^s W_x} \\ 
&=& W_\mu(x)\expf^{-ig\int_0^1 W_x}- 0 
+\int_0^1 ds\, \expf^{-ig\int_s^1 W_x}
 \partial_s X^\nu \partial_\mu X^\rho G_{\rho\nu}  
 \expf^{-ig\int_0^s W_x} \\ 
&=& W_\mu(x)\expf^{-ig\int_0^1 W_x}  
-\int_0^1 ds\, \expf^{-ig\int_s^1 W_x}
 \partial_s X^\nu \partial_\mu X^\rho G_{\nu\rho}  
 \expf^{-ig\int_0^s W_x} \label{cohom-rel01}
\ena

where Eq.~\ref{residue01} and  Eq.~\ref{residue02} make use of 
 Eq.~\ref{expfprop01}. The result 
of Eq.~\ref{cohom-rel01} is nothing but the non-abelian equivalent 
of Eq.~\ref{de-rahm00}, and it can be interesting to 
rewrite it as follows: 

\bea
W_\mu(x) &=& \int_0^1 ds\, \expf^{-ig\int_s^1 W_x}
 \partial_s X^\nu \partial_\mu X^\rho G_{\nu\rho}  
 \expf^{-ig\int_0^s W_x} 
\left(\expf^{-ig\int_0^1 W_x}\right)^{-1} 
+\frac{i}{g}\partial_\mu \Phi \Phi^{-1} 
\ena

This expression gives $W_\mu(x)$ as a gauge equivalent 
of (see Eq.~\ref{gt02}): 

\bea
{W'}_\mu &=& \left(
\expf^{-ig\int_0^1 W_x}\right)^{-1} 
\int_0^1 ds\, \expf^{-ig\int_s^1 W_x}
 \partial_s X^\nu \partial_\mu X^\rho G_{\nu\rho}  
 \expf^{-ig\int_0^s W_x} \nonumber \\ 
&=& \int_0^1 ds\, \left(\expf^{-ig\int_0^s W_x}\right)^{-1} 
 \partial_s X^\nu \partial_\mu X^\rho G_{\nu\rho}  
\left( \expf^{-ig\int_0^s W_x}\right)  
\ena

\end{document}